# 8.7-W average power, in-band pumped femtosecond Ho:CALGO laser at 2.1 µm

**WEICHAO YAO,[1] YICHENG WANG,[1,*] SERGEI TOMILOV,[1] MARTIN HOFFMANN,[1] SHAHWAR AHMED,[1] CHRISTOPH LIEBALD,[2] DANIEL RYTZ,[2] MARK PELTZ,[2] VOLKER WESEMANN,[2] AND CLARA J. SARACENO[1]**

[1]*Photonics and Ultrafast Laser Science, Ruhr Universität Bochum, Universitätsstraße 150, 44801 Bochum, Germany*
[2]*Electro-Optics Technology GmbH, Struthstraße 2, 55743 Idar-Oberstein, Germany*
*\*yicheng.wang@ruhr-uni-bochum.de*

**Abstract:** We report on an in-band pumped SESAM mode-locked Ho:CALGO bulk laser with a record-high average power of 8.7 W and an optical-to-optical efficiency of 38.2% at a central wavelength of 2.1 µm. At this power level, the bulk laser generates pulses with a duration of 369 fs at 84.4-MHz repetition rate, corresponding to a pulse energy of 103 nJ and a peak power of 246 kW. To the best of our knowledge, this is the highest average power and pulse energy directly generated from a mode-locked bulk laser in the 2-3 µm wavelength region. Our current results indicate that Ho:CALGO is a competitive candidate for average power scaling of 2-µm femtosecond lasers.

## 1. Introduction

Power-scaling of ultrafast 2-µm bulk lasers is currently an active topic of research in laser technology, fueled by growing interest in many areas such as spectroscopy [1], material processing [2], and for driving nonlinear conversion to the mid-infrared [3], THz [4], and extreme ultraviolet spectral regions [5]. Currently, most watt-level ultrafast laser systems at this wavelength are based on multi-stage parametric amplifiers pumped by powerful 1-µm lasers, resulting in high complexity and cost [6]. In contrast, laser systems based on gain media directly emitting at 2 µm are an attractive alternative due to their simplicity. In particular, powerful mode-locked oscillators in this spectral region are of critical relevance for applications demanding high repetition rates or for seeding high-power amplifiers. In this context, finding suitable broadband gain materials with high thermal conductivity is an active topic of research, and several $Tm^{3+}$-, $Ho^{3+}$-, and $Cr^{2+}$-doped bulk lasers are being explored, covering the spectral range from 1.9 µm to 2.4 µm.

$Cr^{2+}$-doped gain media (Cr:ZnSe/ZnS) are attractive because of their ultrabroad gain bandwidth, allowing for the generation of short pulses down to few cycles [7-9]. With respect to average power, a 2-W femtosecond Kerr-lens mode-locked Cr:ZnS oscillator has been reported with 67-fs pulse duration [7], which represents thus far the highest average power among mode-locked 2-µm bulk lasers. However, the large thermal optical coefficients of Cr:ZnSe/ZnS have represented a hurdle for scaling oscillator average power to significantly higher values [10]. On the other hand, $Tm^{3+}$- and $Ho^{3+}$-doped gain media offer great potential to scale the average power of ultrafast 2-µm lasers. In fact, continuous-wave (CW) bulk lasers with hundreds of watts of power have been demonstrated with these dopants based on different hosts [11-13]. However, most femtosecond $Tm^{3+}$- or $Ho^{3+}$-doped bulk lasers only achieve average powers of a few hundred milliwatts [14], mostly due to the specific optical and thermal properties of each gain media used and the specific pump configuration available. $Tm^{3+}$-doped materials can be pumped by high-power diodes; however, the low beam quality of these pump sources reduces the optical-to-optical efficiency in the bulk geometry. To date, the highest average power of mode-locked Tm bulk laser is at 1-W level pumped with a 1.6-µm fiber laser [15, 16]. The limited pump power of the 1.6-µm fiber laser has been the limitation to scale their average output power.

In-band pumping of mode-locked $Ho^{3+}$-doped lasers combines the advantages of high-power single-mode Tm-fiber laser [17] pumping and materials with typically high thermal conductivity, thus making it a promising route for achieving high average power and efficient mode-locking at 2 µm. However, the pulse duration of $Ho^{3+}$-doped bulk lasers remained thus far in the few picoseconds range, mostly due to the limited and structured gain bandwidth available from classically used materials such as Ho:YAG [18] and Ho:YLF [19]. In this respect, $Ho^{3+}$-doped disordered $CaLnAlO_4$ (Ln stands for Gd/Y/Lu) crystals, which features high thermal conductivity, weak thermal expansion, and negative thermo-optic coefficients in combination with expected broad and smooth emission spectra are attractive for high average power femtosecond operation [20, 21]. Early measurements reported a 30-nm gain bandwidth for $Ho:CaGdAlO_4$ (Ho:CALGO) crystal [21], which is well-suited for generating femtosecond pulses. Using this family of materials, only CW results based on $Ho:CaYAlO_4$ have been reported [22].

Motivated by the successful results achieved with Yb:CALGO and $Yb:CaYAlO_4$ lasers where over 10 W, sub-100 fs have been demonstrated [23, 24], we explore here high-power CW and mode-locking operation of Ho:CALGO laser. The Ho:CALGO laser was in-band pumped with a single-mode Tm-doped fiber laser pump source and mode-locked with a SESAM, reaching an average output power of 8.7 W and a pulse duration of 369 fs with an optical-to-optical efficiency of 38.2%. To the best of our knowledge, this is the first report on a mode-locked laser with this promising gain material and the highest average power and pulse energy directly generated from any mode-locked femtosecond bulk laser in the 2-3 µm wavelength region.

## 2. Experimental Setup

The crystal used throughout these experiments is a 3.1 at.% doping concentration Ho:CALGO crystal (developed at Electro-Optics Technology GmbH, Germany) grown with the Czochralski method in a boule size of 30 mm (diameter)×87 mm (length). Two crystal samples with one plane cut and one Brewster's angle cut with an effective aperture of 4 mm×4 mm and a length of 15 mm were cut along an *a*-axis from the boule, as shown in Fig. 1. The pump laser is a randomly polarized single-mode Tm-fiber laser providing up to 30 W of CW power at 1940 nm. To measure the pump absorption, a linearly polarized laser was achieved first with a Glan-Taylor polarizer. We measured the small signal absorption coefficients of a plane cut Ho:CALGO crystal at 1940 nm along *E*//*a* and *E*//*c* to be 1.1 $cm^{-1}$ and 1.9 $cm^{-1}$, respectively. For the laser experiment, we used the Brewster's angle cut crystal. It was wrapped with a 0.03-mm indium foil, mounted in a water-cooled copper holder (kept at 16 ºC), and then placed in an astigmatism-compensated resonator shown in Fig. 1. The folding angles of input mirror M1 (anti-reflection, AR-coated at 1940 nm and high-reflection, HR-coated at 2050-2400 nm, 100-mm Radius of curvature) and M2 (HR-coated at 1900-2150 nm, $R = 300$ mm) were ~10º and ~9º, respectively.

For CW experiments, we tested the resonator with mirrors M1, M2, high-reflectivity (HR at 1900-2150 nm) mirror HR2, and different output couplers (OC, $T_{OC} = 1\%, 3\%, 5\%, 7\%, 10\%$, and HR at 1940 nm). A double-pass of the pump light enabled ~80% absorption of the incident pump power, taking into account the Fresnel reflection at the two Brewster surfaces. Approximately 22.8 W was absorbed by the crystal considering the losses of the lens and input mirror. The pump light was focused on the crystal with a beam radius of 93 µm (sagittal) × 175 µm (tangential). The total cavity length is ~1.5 m, with a calculated $TEM_{00}$ laser mode radius in the crystal of 60 µm (sagittal) × 130 µm (tangential), leading to slightly multimode operation.

For mode-locking operation, the total cavity length is ~1.8 m with a $TEM_{00}$ laser mode radius of 95 µm (sagittal) × 185 µm (tangential) in the crystal, which ensures single-mode operation and excellent mode-matching between the pump and mode-locking laser modes. To ensure stable soliton mode-locking [25], in addition to the -1800 $fs^2$ of round-trip group delay dispersion (GDD) from the 15-mm CALGO crystal at 2.1 µm [26], additional four dispersive mirrors DM1-4 with multiple bounces were used to increase the total negative GDD. The total round-trip GDD was then approximately -16800 $fs^2$. A commercially available GaSb-based SESAM (RefleKron Ltd.) with a working range of 2.1-2.2 µm was used to start and stabilize the mode-locking mechanism. We measured the nonlinear reflectivity parameters of the SESAM at the laser wavelength and obtained the following parameters: a saturation fluence ($F_{sat}$) of 10 µJ/$cm^2$, a modulation depth ($\Delta R$) of 0.23%, and a non-saturable loss ($\Delta R_{ns}$) of 0.12%. We present the SESAM measurements in the Results section, with a discussion of the obtained parameters. The calculated GDD of the SESAM in this wavelength region is less than 150 $fs^2$. The beam radius on the SESAM was calculated to be 470 µm.

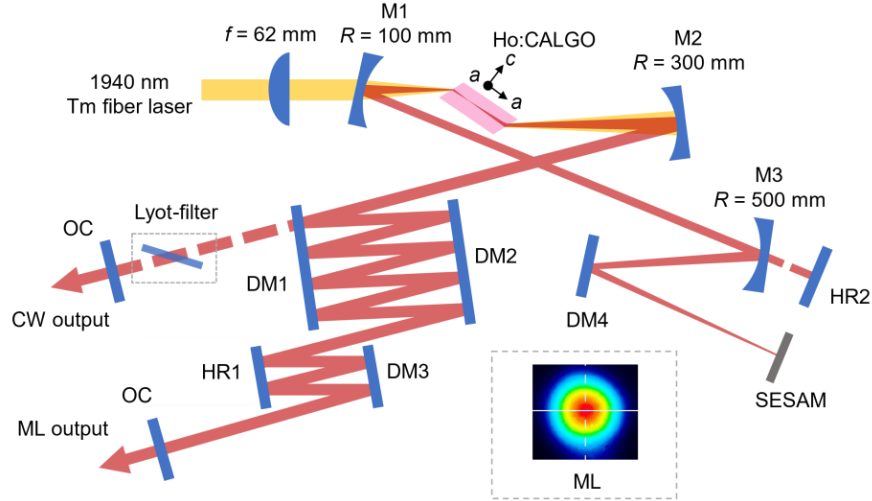

Fig. 1. Experimental setup of SESAM mode-locked Ho:CALGO laser. The arrows next to the crystal indicate the crystal axis. DM1-DM4: dispersion mirrors. DM1, DM3, DM4: -500 $fs^2$ per bounce; DM2: -1000 $fs^2$ per bounce. CW: Continuous wave. ML: Mode-locking. Inset: far-field beam profile of ML at 8.7-W average output power.

## 3. Results and discussions

### 3.1 CW laser operation

We first conducted CW laser experiments with the above described first cavity in order to test the capabilities of the gain material. With this four-mirror cavity, the laser supported slightly multi-mode operation since the laser mode radius was much smaller than the pump radius. Figure 2(a) shows the CW laser output power dependence on absorbed pump power with different $T_{OC}$. The output power was only recorded at a low pump power (<8 W) when $T_{OC}$ = 1% due to a low slope efficiency ($\eta$ = 32.7%), and the final output power was 2 W. The slope efficiency increases to reach 47.3%, 50.5%, 52.3%, and 52.4% with $T_{OC}$ = 3%, 5%, 7%, 10%, respectively. A maximum output power of 10.8 W was obtained at 22.8 W of absorbed pump power with $T_{OC}$ = 7%, with an optical-to-optical efficiency of 47.4%. A roll over of optical-to-optical efficiency was not observed with four OCs even at the highest absorbed pump power, as shown in Fig. 2(b), illustrating the excellent thermal properties of the laser crystal. We used the knife-edge method to measure the beam quality at 10.8-W output power. The $M^2$ was fitted to be 1.27 and 1.55 in sagittal and tangential directions, respectively. The slight difference of the beam quality in two directions can be attributed to the anisotropic property of Ho:CALGO crystal and weak astigmatism in the four-mirror cavity.

The laser spectra with different $T_{OC}$ were recorded by a WaveScan spectrometer (A.P.E. GmbH) with a spectral resolution of 0.5 nm, as shown in Fig. 2(c). The laser wavelength shows a significant red-shift from 2121 nm to 2135 nm when $T_{OC}$ decreases from 10% to 1%. This is due to the enhanced reabsorption effect of the $Ho^{3+}$ quasi-three-level system when $T_{OC}$ becomes lower. Thereafter, a Lyot-filter (5-mm thick quartz birefringent plate) was inserted into the resonator at a Brewster angle to measure the tuning bandwidth of the material. The laser wavelength was tuned over 80 nm from 2068 nm to 2148 nm by rotating the filter when $T_{OC}$ = 5%, as shown in Fig. 2(d), whereas the longer wavelength output was limited by the free spectral range of the filter. Note that this operation was performed at 4.5 W of absorbed pump power to reduce the risk of crystal damage, but the tuning range should not change significantly at a higher pump power since the gain spectrum is clamped by the cavity loss. However, it can be predicted that the tuning range will increase with a lower $T_{OC}$ because of a wider gain spectrum. Based on the measured tuning range, achieving sub-100 fs pulse duration appears possible from a mode-locked Ho:CALGO laser, even at higher inversion levels.

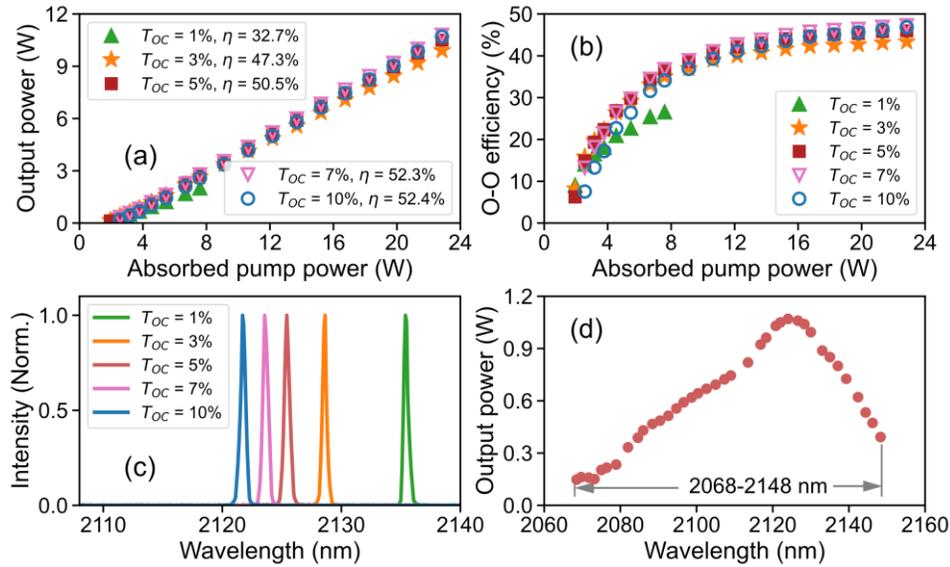

Fig. 2. CW performance of Ho:CALGO laser. (a) Output power dependence on absorbed pump power with different $T_{OC}$; (b) Optical-to-optical efficiency; (c) Measured wavelength with different $T_{OC}$; (d) Wavelength-tunability with $T_{OC}$ = 5% when the absorbed pump power was 4.5 W.

### *3.2 Mode-locking experiments*

For the mode-locking experiments, we used a slightly longer resonator and replaced SESAM as a HR mirror allowing us to obtain fundamental transverse mode CW operation as described in more detail in section 2. In this case, the CW output powers at 22.8 W of absorbed pump power with $T_{OC}$ = 7% and $T_{OC}$ = 10% were only slightly lower than the experiments described in the previous section 3.1, with 9.6 W and 9.4 W, respectively. This included the intracavity dispersive mirrors intracavity that were then used for the mode-locking experiments. The $M^2$ at 9.6-W output power was measured to be 1.04, confirming fundamental mode operation.

For stable soliton mode-locking, we replaced the end HR mirror by the above-mentioned commercially available GaSb-SESAM and compared the laser performance with different $T_{OC}$. Stable mode-locking was achieved in all cases, naturally reaching different maximum output power levels because of soliton mode-locking criteria, after which we observed mode-locking instability which in our case started with the appearance of CW spikes in the laser spectrum. In stable mode-locked operation, the results at the highest average output powers are summarized in Fig. 3 and Table 1. With $T_{OC} = 1\%$, the measured spectral bandwidth ($\Delta\lambda$, full-width at half maximum) was 15.1 nm at a center wavelength of 2133.8 nm. The pulse duration was measured with a noncollinear autocorrelator (pulseCheck, A.P.E. GmbH) and results are shown in Fig. 3(b). Fitting with a sech$^2$-function gives a pulse duration ($\Delta\tau$, full-width at half maximum) of 306 fs, corresponding to the Fourier limit of the measured spectrum. The calculated time-bandwidth product (TBP) was 0.304. A slightly smaller TBP than the Fourier limited value (0.315) can be attributed to the slight deviation of actual laser pulses from perfect soliton-shaped pulses. The pulse duration was 369 fs when 7% $T_{OC}$ was used, with a spectral bandwidth of 12.4 nm at a center wavelength of 2121.7 nm. The TBP was calculated to be 0.305. A shorter pulse duration with a lower $T_{OC}$ is possibly because of a larger and flatter gain bandwidth, this allows the laser to generate a shorter pulse duration with a low modulation depth [27, 28]. In addition, a slightly lower negative GDD also has a slight contribution to the shorter pulse duration with 1% $T_{OC}$, as shown in Fig. 3(a).

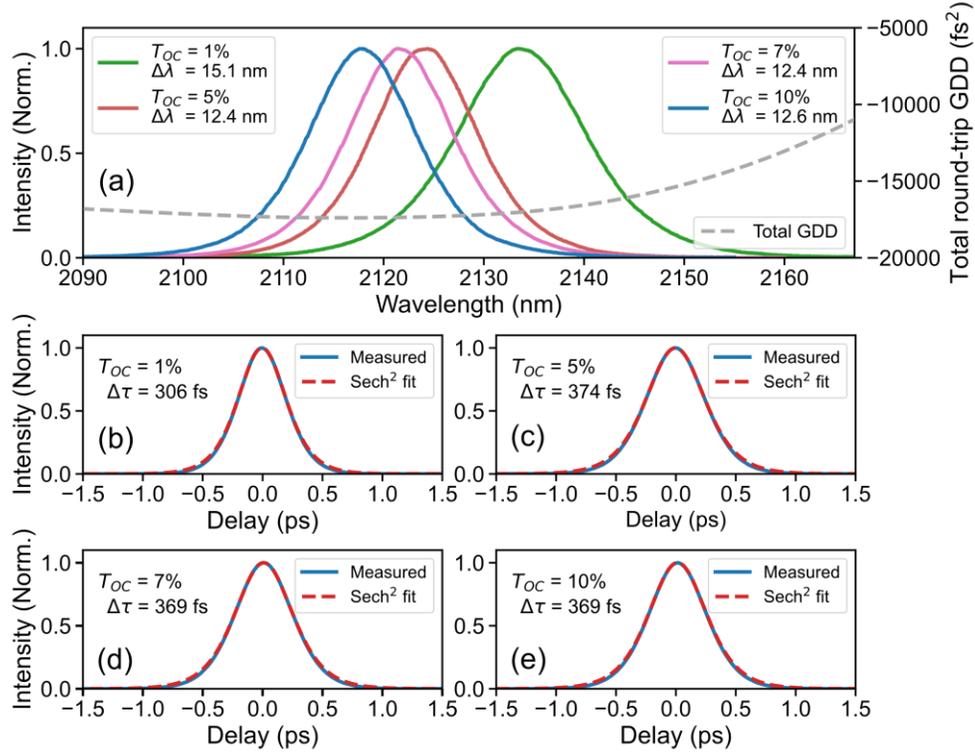

Fig. 3. Mode-locked laser characters at the highest average output power with different $T_{OC}$. (a) Laser spectra; (b)-(e) Noncollinear autocorrelation traces.

Table 1. Mode-locked Ho:CALGO laser with -16800 fs$^2$ of round-trip GDD. $P_{ab}$: the maximum absorbed pump power in stable mode-locking. $P_{out}$: average output power.

| $T_{OC}$ (%) | $P_{ab}$ (W) | $P_{out}$ (W) | $\lambda$ (nm) | $\Delta\lambda$ (nm) | $\Delta\tau$ (fs) | TBP |
|---|---|---|---|---|---|---|
| 1 | 7.2 | 1 | 2133.8 | 15.1 | 306 | 0.304 |
| 5 | 12.5 | 4.1 | 2124.1 | 12.4 | 374 | 0.308 |
| 7 | 16.7 | 6.1 | 2121.7 | 12.4 | 369 | 0.305 |
| 10 | 22.8 | 8.7 | 2117.9 | 12.6 | 369 | 0.311 |

With $T_{OC}$ = 10%, mode-locking started at an absorbed pump power of 20.2 W, and the average output power was scaled to 8.7 W at the maximum absorbed pump power, i.e., 22.8 W, corresponding to an optical-to-optical efficiency of 38.2%. During the pump power increase while mode-locking was stable from 20.2 W to 22.8 W, the optical-to-optical efficiency increased continuously. This seems to indicate that thermal effects will not be a limit to scale the average output power further when more pump power is available. At 8.7-W output power, the spectral bandwidth was measured to be 12.6 nm at a center wavelength of 2117.9 nm, as shown in Fig. 3(a). A pulse duration of 369 fs gives a TBP of 0.311. The far field beam pattern measured with a camera at 8.7 W of average output power is shown in the inset of Fig. 1. The radii ratio of 1:1.01 in sagittal and tangential directions means astigmatism has been completely compensated for.

The stability of mode-locking at the maximum output power was characterized with a 12.5-GHz photodiode and a 1-GHz radio frequency analyzer. In Fig. 4(a), the beat notes of almost same intensity in a 1-GHz scanning range indicate stable mode-locking without modulation. The repetition rate was measured to be 84.4 MHz with a signal-to-noise ratio of 52 dB, resulting in a maximum pulse energy of 103 nJ and a peak power of 246 kW in our experiments. Figure 4(c) shows the mode-locking pulses measured with the 12.5-GHz photodiode and a 25-GHz sampling oscilloscope (PicoScope 9000, Pico Tech.). The time delay between two pulses was ~11.8 ns, which was consistent with the frequency spacing of the resonator, confirming single pulse mode-locking in our experiment.

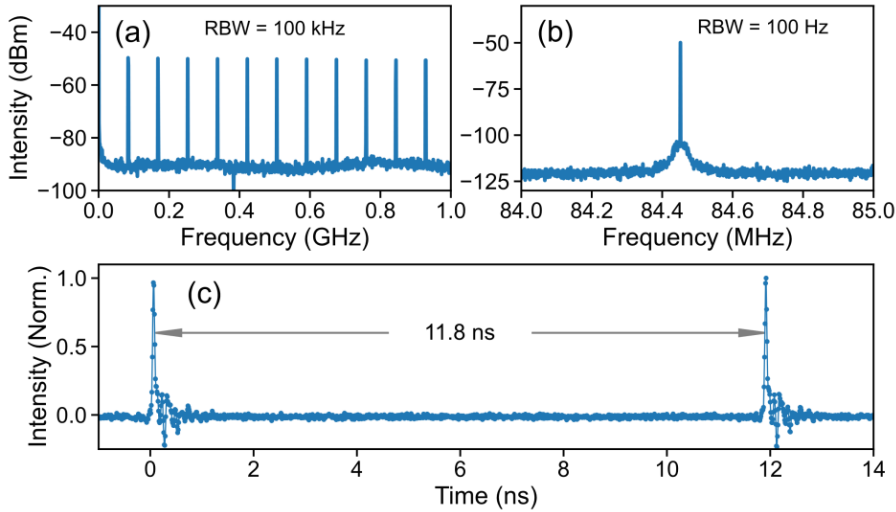

Fig. 4. Radio frequency spectra in (a) 1 GHz and (b) 1 MHz scanning range; (c) Sampling oscilloscope measurement result. RBW: resolution bandwidth.

*3.3 Discussion on current limitations*

In order to understand the limitations of our laser in terms of pulse duration and average power, it is critical to evaluate the nonlinear reflectivity of the SESAM at the operation wavelength and with a comparable pulse duration to that of the laser. The commercially available SESAM did not provide these parameters, therefore we performed our own measurements, using the presented mode-locked laser as a seed for a home-built nonlinear reflectivity setup.

The setup was built following the guidelines presented in [29] and seeded by our 2130-nm mode-locked 350-fs Ho:CALGO laser. We used an input power of 2 W into the setup and focused on the SESAM with a maximum incident fluence of 230 µJ/cm². Figure 5 shows the measured nonlinear reflectivity dependence on the incident fluence. Using an appropriate fitting function, we extract the following saturation fluence ($F_{sat}$) of 10 µJ/cm², a modulation depth ($\Delta R$) of 0.23%, and a non-saturable loss ($\Delta R_{ns}$) of 0.12%. The excellent measurement accuracy of such a small modulation depth was made possible by the excellent long-term stability of the mode-locked laser.

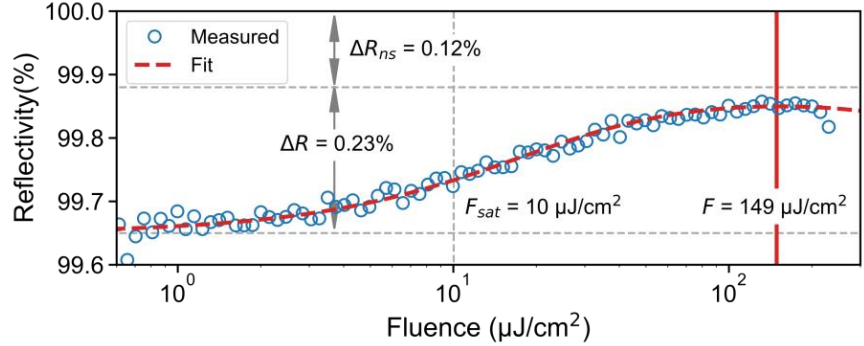

Fig. 5. Nonlinear reflectivity of the SESAM. The solid red line marks the operation fluence at the highest average power in our experiment.

In our mode-locking experiments, the minimum pulse durations achievable with all OCs were in a comparable region, which is a sign that the gain bandwidth is not a limiting factor for shorter pulses. The low modulation depth of 0.23% is then the main limiting factor towards reaching shorter pulse durations in our experiments. In fact, trying to reduce the pulse duration by reducing the intracavity GDD - even at the lowest $T_{OC}$ of 1% - was not able to sustain stable mode-locking with much shorter pulses. Only the shortest value of 279 fs was achieved with a lower GDD of -14800 fs$^2$, at 0.97 W of average output power. This is a clear indication that the limitation to reach shorter pulses is the low modulation depth of the SESAM, and this is confirmed by the observed appearance of CW spikes. In fact, in soliton mode-locking, a lower effective gain cross-section in mode-locking operation compared to CW operation will make the suppression of CW background difficult with moderate modulation depth [27, 28], when the pulse duration becomes short. Such CW spikes can be suppressed by using a SESAM with a higher modulation depth [27, 28]. We believe this should be straightforward to achieve, as the parameters of this SESAM were not designed for our specific wavelength of operation. Furthermore, a straightforward path to obtain shorter pulses is to use Kerr-lens mode-locking, this can be achieved by using a new resonator [30].

Further, in our experiment, the maximum fluence on the SESAM was estimated to be 149 µJ/cm$^2$. This value is at the point of maximum reflectivity, as shown in Fig. 5, which allows to reach high output power levels. Note that there is no significant drop in reflectivity even for the fluence of 200 µJ/cm$^2$, therefore, further scaling of output power in the current scheme to 10 W level should be straightforward, if using higher pump power. An optimized SESAM design with shifted rollover [31], would even allow to operate with small spot sizes at higher saturation values.

## 4. Conclusion and outlook

In conclusion, we present the first in-band pumped high power CW and mode-locking operation of a Ho:CALGO bulk laser. In CW operation, 10.8 W of output power was achieved with a slope efficiency of 52.3% and an optical-to-optical efficiency of 47.3%. We present broadband tunability with a Lyot-filter, showing a tuning range of 80 nm from 2068 nm to 2148 nm. Best SESAM mode-locked performance achieved was 8.7 W of average power, 103 nJ of pulse energy, and 369 fs of pulse duration. Benefiting from the high optical-to-optical efficiency (38.2%) of an in-band pumping scheme, our current results represent the highest average power and pulse energy to date achieved in the bulk geometry in the 2-3 µm wavelength region.

Additionally, we identify the main limiting factors in our experiment towards reaching shorter pulse durations at this average power level to be the low modulation depth of our SESAM, which we characterized in detail using a home-built setup. In the near future, we believe over 10 W of average power with sub-100 fs pulse duration should be feasible using SESAMs with higher modulation depth or Kerr-lens mode-locking, in combination with a higher absorbed pump power (for example using higher doping crystals and/or a higher incident pump power).

**Funding.** European Research Council (805202); Deutsche Forschungsgemeinschaft (390677874).

**Acknowledgment.** These results are part of a project that has received funding from the European Research Council (ERC) under the European Union's Horizon 2020 research and innovation programme (grant agreement No. 805202 - Project Teraqua). Funded by the Deutsche Forschungsgemeinschaft (DFG, German Research Foundation) under Germanys Excellence Strategy – EXC-2033 – Projektnummer 390677874 – RESOLV. We acknowledge support by the Open Access Publication Funds of the Ruhr-Universität Bochum. W. Yao acknowledges financial support from the Alexander von Humboldt Foundation through a Humboldt Research Fellowship.

**Disclosures.** The authors declare no conflicts of interest.

**Data availability.** Data underlying the results presented in this paper are not publicly available at this time but may be obtained from the authors upon reasonable request.

## References

1. J. Mandon, G. Guelachvili, N. Picqué, "Fourier transform spectroscopy with a laser frequency comb," Nature Photon **3**, 99–102 (2009).
2. R. R. Gattass, E. Mazur, "Femtosecond laser micromachining in transparent materials," Nature Photon **2**, 219–225 (2008).
3. N. Leindecker, A. Marandi, R. L. Byer, K. L. Vodopyanov, J. Jiang, I. Hartl, M. Fermann, and P. G. Schunemann, "Octave-spanning ultrafast OPO with 2.6-6.1 µm instantaneous bandwidth pumped by femtosecond Tm-fiber laser," Opt. Express **20**(7), 7046-7053(2012).
4. M. Clerici, M. Peccianti, B. E. Schmidt, L. Caspani, M. Shalaby, M. Giguère, A. Lotti, A. Couairon, F. Légaré, T. Ozaki, D. Faccio, and R. Morandotti, "Wavelength scaling of terahertz generation by gas ionization," Phys. Rev. Lett. **110**, 253901(2013).
5. T. Popmintchev, M. C. Chen, D. Popmintchev, P. Arpin, S. Brown, S. Ališauskas, G. Andriukaitis, T. Balčiunas, O. Mücke, A. Pugzlys, A. Baltuška, B. Shim, S. E. Schrauth, A. Gaeta, C. H. García, L. Plaja, A. Becker, A. J. Becker, M. M. Murnane, H. C. Kapteyn, "Bright coherent ultrahigh harmonics in the keV x-ray regime from mid-infrared femtosecond lasers," Science, **336**(6086): 1287-1291(2012).
6. T. Feng, A. Heilmann, M. Bock, L. Ehrentraut, T. Witting, H. Yu, H. Stiel, S. Eisebitt, and M. Schnürer, "27 W 2.1 µm OPCPA system for coherent soft X-ray generation operating at 10 kHz," Opt. Express **28**(6), 8724-8733(2020).
7. S. B. Mirov, V. V. Fedorov, D. Martyshkin, I. S. Moskalev, M. Mirov, and S. Vasilyev, "Progress in mid-IR lasers based on Cr and Fe-doped II-IV chalcogenides," IEEE J. Sel. Top. Quantum Electron. **21**(1), 1601719(2015).
8. N. Nagl, S. Gröbmeyer, V. Pervak, F. Krausz, O. Pronin, and K. F. Mak, "Directly diode-pumped, Kerr-lens mode-locked, few-cycle Cr:ZnSe oscillator," Opt. Express **27**(17), 24445-24454(2019).
9. S. Vasilyev, I. Moskalev, V. Smolski, J. Perrers, M. Mirov, Y. Barnakov, V. Fedorov, D. Martyshkin, S. Mirov, and V. Gapontsev, "Kerr-lens mode-locked Cr:ZnS," Opt. Express **29**(2), 2458-2465(2021).
10. I. Moskalev, S. Mirov, M. Mirov, S. Vasilyev, V. Smolski, A. Zakrevskiy, and V. Gapontsev, "140 W Cr:ZnSe laser system," Opt. Express **24**(18), 21090-21104(2016).
11. P. Liu, L. Jin, X. Liu, H. Huang, and D. Shen, "High efficiency Tm:YAG slab laser with hundred-watts-level output power," Appl. Opt. **55**(10), 2498–2502 (2016).
12. W. C. Yao, E. H. Li, Y. J. Shen, C. Y. Ren, Y. G. Zhao, D. Y. Tang and D. Y. Shen, "A 142 W Ho:YAG laser single-end-pumped by a Tm-doped fiber laser at 1931 nm," Laser Phys. Lett. **16**, 115001(2019).
13. F. Wang, J. Tang, E. Li, C. Shen, J. Wang, D. Tang, and D. Shen, "$Ho^{3+}$:$Y_2O_3$ ceramic laser generated over 113 W of output power at 2117 nm," Opt. Lett. **44**(24), 5933-5936(2019).
14. J. Ma, Z. Qin, G. Xie, L. Qian, and D. Tang, "Review of mid-infrared mode-locked laser sources in the 2.0 µm-3.5 µm spectral region," Appl. Phys. Rev. **6**, 021317(2019).
15. M. Tokurakawa, E. Fujita, and C. Kränkel, "Kerr-lens mode-locked $Tm^{3+}$:$Sc_2O_3$ single-crystal laser in-band pumped by an Er:Yb fiber MOPA at 1611 nm," Opt. Lett. **42**(16), 3185-3188(2017).
16. N. Zhang, Z. Wang, S. Liu, W. Jing, H. Huang, Z. Huang, K. Tian, Z. Yang, Y. Zhao, U. Griebner, V. Petrov, and W. Chen, "Watt-level femtosecond Tm-doped "mixed" sesquioxide ceramic laser in-band pumped by a Raman fiber laser at 1627 nm," Opt. Express **30**(13), 23978-23985(2022).
17. W. Yao, C. Shen, Z. Shao, J. Wang, F. Wang, Y. Zhao, and D. Shen, "790 W incoherent beam combination of a Tm-doped fiber laser at 1941 nm using a 3×1 signal combiner," Appl. Opt. **57**(20), 5574-5577(2018).

18. Y. Wang, R. Lan, X. Mateos, J. Li, C. Hu, C. Li, S. Suomalainen, A. Härkönen, M. Guina, V. Petrov, and U. Griebner, "Broadly tunable mode-locked Ho:YAG ceramic laser around 2.1 μm," Opt. Express **24**(16), 18003-18012(2016).
19. N. Coluccelli, A. Lagatsky, A. Di Lieto, M. Tonelli, G. Galzerano, W. Sibbett, and P. Laporta, "Passive mode locking of an in-band-pumped Ho:YLiF$_4$ laser at 2.06 μm," Opt. Lett. **36**(16), 3209-3211(2011).
20. Z. Pan, P. Loiko, S. Slimi, H. Yuan, Y. Wang, Y. Zhao, P. Camy, E. Dunina, A. Kornienko, L. Fomicheva, L. Wang, W. Chen, U. Griebner, V. Petrov, R. M. Solé, F. Díaz, M. Aguiló, X. Mateos, "Tm,Ho:Ca(Gd,Lu)AlO$_4$ crystals: Crystal growth, structure refinement and Judd-Ofelt analysis," J. Lumin. **246**, 118828(2022).
21. Y. Wang, P. Loiko, Y. Zhao, Z. Pan, W. Chen, M. Mero, X. Xu, J. Xu, X. Mateos, A. Major, M. Guina, V. Petrov, and U. Griebner, "Polarized spectroscopy and SESAM mode-locking of Tm,Ho:CALGO," Opt. Express **30**(5), 7883-7893(2022).
22. D. Zhou, J. Di, C. Xia, X. Xu, F. Wu, J. Xu, D. Shen, T. Zhao, A. Strzęp, W. Ryba-Romanowski, and R. Lisiecki, "Spectroscopy and laser operation of Ho:CaYAlO$_4$," Opt. Mat. **3**(3), 339-345(2013).
23. W. Tian, R. Xu, L. Zheng, X. Tian, D. Zhang, X. Xu, J. Zhu, J. Xu, and Z. Wei, "10-W-scale Kerr-lens mode-locked Yb:CALYO laser with sub-100-fs pulses," Opt. Lett. **46**(6), 1297-1300(2021).
24. A. Greborio, A. Guandalini, and J. Aus der Au, "Sub-100 fs pulses with 12.5 W from Yb:CALGO based oscillators," Solid State Lasers XXI: Technology and Devices, in SPIE Photonics West 2012, paper 823511.
25. F. X. Kärtner, I. D. Jung, and U. Keller, "Soliton mode-locking with saturable absorber," IEEE J. Sel. Top. Quantum Electron. **2**(3), 540-556(1996).
26. P. Loiko, P. Becker, L. Bohatý, C. Liebald, M. Peltz, S. Vernay, D. Rytz, J. M. Serres, X. Mateos, Y. Wang, X. Xu, J. Xu, A. Major, A. Baranov, U. Griebner, and V. Petrov, "Sellmeier equations, group velocity dispersion, and thermo-optic dispersion formulas for CaLnAlO$_4$ (Ln=Y,Gd) laser host crystals," Opt. Lett. **42**(12), 2275-2278(2017).
27. I. J. Graumann, A. Diebold, C. G. E. Alfieri, F. Emaury, B. Deppe, M. Golling, D. Bauer, D. Sutter, C. Kränkel, C. J. Saraceno, C. R. Phillips, and U. Keller, "Peak-power scaling of femtosecond Yb:Lu$_2$O$_3$ thin-disk lasers," Opt. Express **25**(19), 22519-22536(2017).
28. M. Tokurakawa, A. Shirakawa, and K. Ueda, "Estimation of Gain Bandwidth Limitation of Short Pulse Duration Based on Competition of Gain Saturation," in Advanced Solid-State Photonics, OSA Technical Digest Series (CD) (Optical Society of America, 2010), paper AMB16.
29. D. J. C. Mass, B. Rudin, A.-R. Bellancourt. D. Iwaniuk, S. V. Marchese, T. Südmeyer, and U. Keller, "High precision optical characterization of semiconductor saturable absorber mirrors," Opt. Express **16**(10), 7571-7579(2008).
30. L. Wang, W. Chen, Y. Zhao, Y. Wang, Z. Pan, H. Lin, G. Zhang, L. Zhang, Z. Lin, J. E. Bae, T. G. Park, F. Rotermund, P. Loiko, X. Mateos, M. Mero, U. Griebner, and V. Petrov, "Single-walled carbon-nanotube saturable absorber assisted Kerr-lens mode-locked Tm:MgWO$_4$ laser," Opt. Lett. **45**(22), 6142-6145(2020).
31. C. J. Saraceno, C. Schriber, M. Mangold, M. Hoffmann, O. H. Heckl, C. R. E. Baer, M. Golling, T. Südmeyer, and U. Keller, "SESAMs for high-power oscillators: design guidelines and damage thresholds," IEEE J. Sel. Top. Quantum. Electron. **18**(1), 29-41(2012).